# Optimization of light collection from crystal scintillators for cryogenic experiments

F.A. Danevich [a], R.V. Kobychev [a,b], V.V. Kobychev [a,c], H. Kraus [d], V.B. Mikhailik [d,e], V.M. Mokina [a]

[a] *Institute for Nuclear Research, MSP 03680, Kyiv, Ukraine*
[b] *National Technical University of Ukraine "Kyiv Polytechnic Institute", 03056 Kyiv, Ukraine*
[c] *Kyungpook National University, Daegu,702-701, Republic of Korea*
[d] *Department of Physics, University of Oxford, Keble Road, Oxford OX1 3RH, UK*
[e] *Diamond Light Source, Harwell Science Campus, Didcot, OX11 0DE, UK*

**ABSTRACT**

High light collection efficiency is an important requirement in any application of scintillation detectors. The purpose of this study is to investigate the possibility for improving this parameter in cryogenic scintillation bolometers, which can be considered as a promising detectors in experiments investigating neutrinoless double beta decay and dark matter. Energy resolutions and relative pulse amplitudes of scintillation detectors using ZnWO$_4$ scintillation crystals of different shapes (cylinder $\varnothing$ 20 × 20 mm and hexagonal prism with diagonal 20 mm and height 20 mm), reflector materials and shapes, optical contact and surface properties (polished and diffused) were measured at room temperature. Propagation of optical photons in these experimental conditions was simulated using Geant4 and ZEMAX codes. The results of the simulations are found to be in good agreement with each other and with direct measurements of the crystals. This could be applied to optimize the geometry of scintillation detectors used in the cryogenic experiments.

## 1. Introduction

Cryogenic scintillators are a promising detector component to search for dark matter and neutrinoless double beta (2β) decay due to excellent energy resolution, particle discrimination ability, and low energy threshold [1, 2]. They also offer the important possibility of using compounds with a choice of different nuclei. For instance, the CRESST collaboration already uses low-temperature CaWO$_4$ scintillating bolometers to search for weakly interacting massive particles (WIMP) [3, 4] while several experiments are in preparation to search for dark matter and 2β decay. In particular, the goal of the EURECA project [5] is to build a ton-scale cryogenic detector to investigate WIMP-nucleon scattering. The aim of a number of R&D projects is to build 2β decay experiments for the exploration of neutrino mass hierarchy scenarios using CdWO$_4$ [6,7], CaMoO$_4$ [8], ZnSe [9, 10, 11], and ZnMoO$_4$ [12, 13, 14] crystal scintillators.

The design and optimization of detectors for the next generation of cryogenic experiments require not only knowledge on the physics of conversion of high-energy quanta to visible photons but also an understanding of the mechanism of light transport in the crystal. Given that less than half of the photons generated in a scintillation crystal reach the photodetector [15, 16, 17] it is apparent that the efficiency of light collection has a significant effect on the overall performance of a detection module. Maximising the light collection is of particular importance for dark matter experiments where a low energy threshold is needed for separation of nuclear recoil events (effect) from gamma/beta events (background) as this eventually defines the experiment's sensitivity [3, 4]. In 2β experiments the light collection is crucial for achieving effective pulse-shape discrimination, in particular for random coincidence events, which were recently recognized as one of the problematic sources of background in bolometric detectors, especially when the search involves the isotope $^{100}$Mo that has a particularly fast two neutrino double beta decay rate [18, 19]. When designing detection modules for such applications it is important to understand and control the factors that influence overall light collection. Therefore, the aim of this work is to investigate how the light collection of scintillation detectors is dependent on the particular choice of the experimental setup, i.e. scintillator shape, surface condition, wrapping, optical contact and reflector.

In this work we studied the performance of ZnWO$_4$ scintillation detectors using experimental and simulation approaches. The choice of the scintillator was motivated by the fact that it is a promising target for dark matter and/or 2β decay experiments due to high light output, very low level of radioactive contamination and its composition (presence of zinc and tungsten nuclei containing potentially 2β active isotopes) [20, 21, 22, 23, 24, 25]. Moreover zinc tungstate has optical properties very similar to those of other representatives of the ABO$_4$ (A = Ca, Zn, Cd, B = Mo, W) family of heavy inorganic scintillators which are considered as attractive complementary targets for cryogenic rare event searches [26]. Therefore, the results on the optimisation of the light collection efficiency for this material can be applicable to other cryogenic scintillating bolometers. We measured the dependence of energy resolution and relative pulse amplitude of ZnWO$_4$ scintillation detectors on the crystal shape (hexagonal and cylindrical), design and material of a reflector, type of optical contact with the photodetector, and optical condition of the crystal scintillator surface (polished and diffuse). These results are discussed in the first part of this paper. In the second part we present results of modelling. Two different Monte Carlo techniques were used to simulate the light transport in the scintillator-detector assembly: Geant4 and ZEMAX. Such an approach offered the advantage for cross-examination of the modelling results and their validation – all too often a stumbling-block of simulations.

## 2. Materials, measurements and results

Two ZnWO$_4$ crystal scintillators were produced from one ZnWO$_4$ crystal ingot in the Institute of Scintillation Materials (Kharkiv, Ukraine). This was done to ensure as much as possible identical optical properties of the two samples used in the tests. The first crystal was of cylindrical shape with diameter 20 mm and height 20 mm. The second crystal was in the form of a hexagonal prism with the larger diagonal 20 mm and height 20 mm. The measurements of the transmittance spectra using a spectrophotometer Shimadzu, UV-3600 evidence that both samples have very similar optical properties (see Fig. 1).

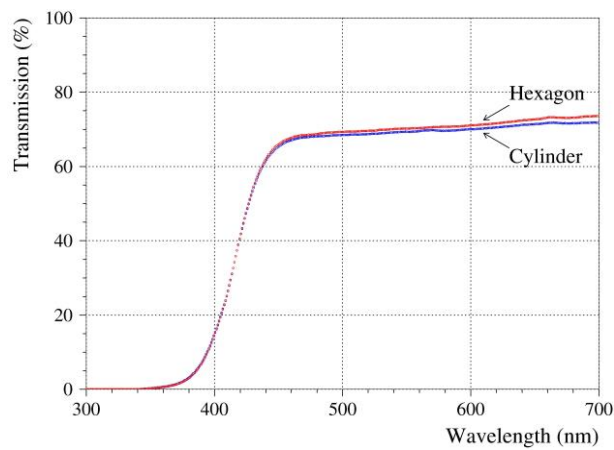

Fig. 1. (Colour online) The optical transmission spectra of cylindrical and hexagonal ZnWO$_4$ crystals measured along the vertical axis.

The measurements of the energy resolution and the relative pulse amplitude with radioactive sources were carried out using a 3" photomultiplier (PMT) Philips XP2412. The positions of the γ sources ($^{137}$Cs and $^{207}$Bi) were chosen to provide a counting rate less than 250 counts/s to avoid overlap of scintillation events due to the rather slow scintillation response of ZnWO$_4$. The energy spectra were accumulated for 40 min, each. Before taking data the experimental setup was allowed to stabilise for 30 minutes after switching on the high voltage of the PMT. The temperature during the measurements was in the range of 21 – 26 °C. Variation of the ZnWO$_4$ light output was estimated to be about 3 % in this temperature interval. To correct for this effect as well as for possible drift in stability the detection system repeated, periodic checks were made through measuring the γ spectrum of a $^{137}$Cs source, using a control ZnWO$_4$ scintillation sample of size 10×10×5 mm$^3$. All experimental data were then processed offline, taking into account the position of the $^{137}$Cs peak measured for the 10×10×5 mm$^3$ control scintillator.

The measurements were carried out for the following arrangements of experimental setups (see Fig. 2):

*A*) ZnWO$_4$ crystal wrapped in 3 layers of PTFE tape and optically coupled to the PMT;

*B*) ZnWO$_4$ crystal surrounded by a cylindrical 3M reflector ⌀ 26 × 25 mm, not in contact with the crystal, and optically coupled to the PMT;

*C*) ZnWO$_4$ crystal surrounded by a cylindrical 3M reflector ⌀ 26 × 25 mm and placed on small acrylic supports (three cubes with dimensions 2 × 2 × 2 mm$^3$) between the crystals and the PMT.

The optical contact in geometries *A* and *B* was provided by Dow Corning Q2-3067 optical gel. It should be noted that geometry *C* represents, to some extent, the conditions of light collection in a cryogenic scintillating detector. There the scintillation crystal must be separated from the other parts of the detector to minimise loss of phonons [22] and not to introduce excess heat capacity.

The measurements were carried out for four conditions of the crystals' surfaces:
1) all surfaces of the ZnWO$_4$ crystal scintillator polished;
2) the side surfaces of the crystals diffuse, the top and bottom surfaces polished;
3) the side surfaces and the top face of the crystals diffuse, the face viewed by the PMT polished;
4) all surfaces of the crystals diffuse.

Lapping of the crystal surfaces was done using sandpaper P1000 (KWH Mirka Ltd) with grain size 18 ± 1 micrometres. The roughness of the diffuse surfaces, evaluated by using an optical microscope, is about 5 – 40 micrometres (see Fig. 3). The roughness of the polished surfaces is about 0.2 micrometres.

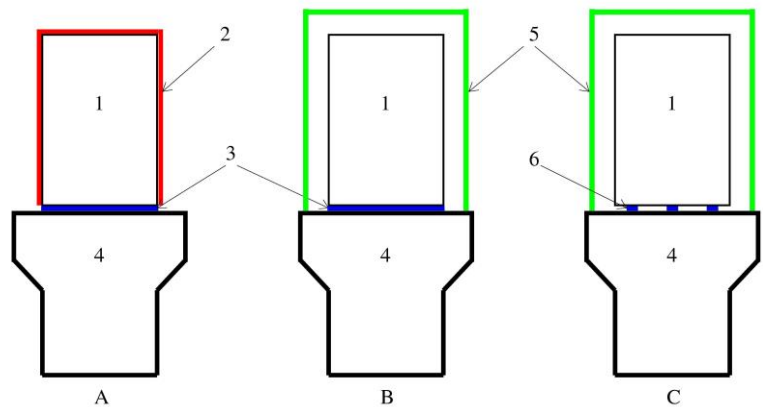

Fig. 2. (Colour online) Schematics of the experimental setups used for the measurements of ZnWO$_4$ crystal scintillators (1 – crystal, 2 – 3 layers of PTFE tape, 3 – optical contact, 4 – PMT, 5 – 3M reflector, 6 – acrylic support).

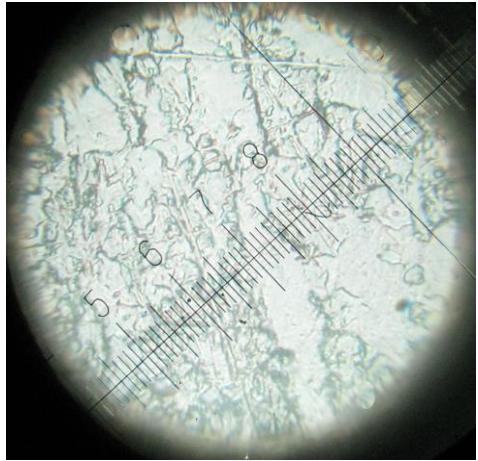

Fig. 3. (Colour online) Micrograph of the ZnWO$_4$ diffuse surface. The small division of the scale is 3.7 micron.

The measurement for surface conditions 1 − 3 was repeated three times and for condition 4 was carried out once.

The energy spectra of $^{137}$Cs and $^{207}$Bi γ quanta presented in Fig. 4 (upper part) were collected for the hexagonal ZnWO$_4$ crystal scintillator with all diffuse surfaces in optical contact with the PMT surrounded by the 3M reflector.

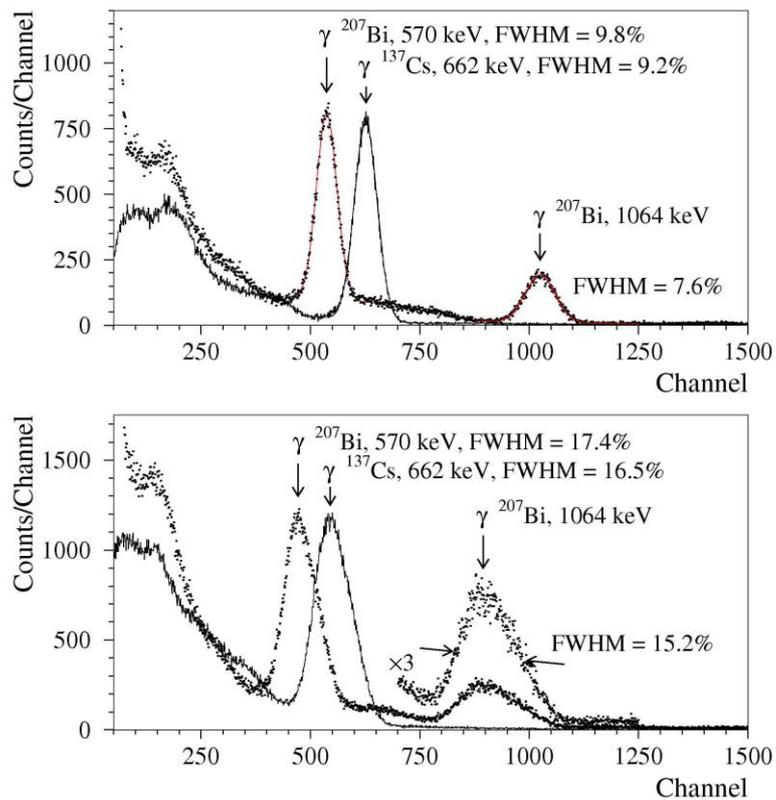

Fig. 4. (Colour online) Energy spectra of $^{137}$Cs and $^{207}$Bi γ rays measured for the ZnWO$_4$ crystal of hexagonal shape with all surface diffuse, in geometry *B* (upper part, see text and Fig. 2). Energy spectra of $^{137}$Cs and $^{207}$Bi γ rays measured for the ZnWO$_4$ crystal of cylindrical shape with all surface polished, in geometry *B* (lower part). The energy resolution with the polished crystal is substantially worse due to the non-uniformity of light collection.

Data on the energy resolution measured with 662 keV γ quanta of $^{137}$Cs for different conditions of experiments are displayed in Fig. 5. It is immediately seen that a cylindrical scintillator has worst energy resolution when compared with a scintillator of hexagonal shape.

The difference in energy resolution is very noticeable for polished surfaces in geometry C and reduces when crystal surfaces are diffused.

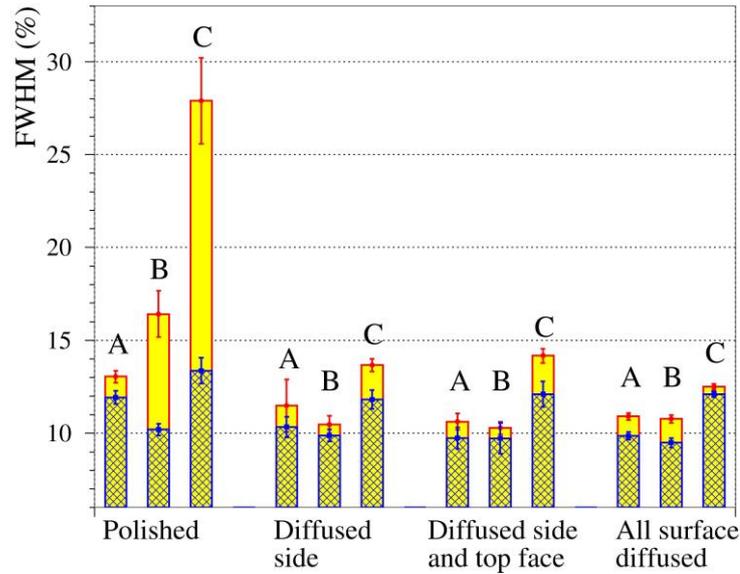

Fig. 5. (Colour online) Energy resolution (full width at half maximum, FWHM) for hexagonal (shaded bars) and cylindrical (clear bars) ZnWO$_4$ scintillator crystals for setups *A, B* and *C* (see Fig. 2) measured with a $^{137}$Cs source for the different conditions of the crystal surface.

Data on the relative pulse amplitude measured with 662 keV γ quanta of $^{137}$Cs for different conditions of experiments are presented in Fig. 6. The experimental data are normalized with respect to the values of light collection efficiency calculated using ZEMAX (see subsection 3.2) for the configuration − geometry *B* − chosen as a reference.

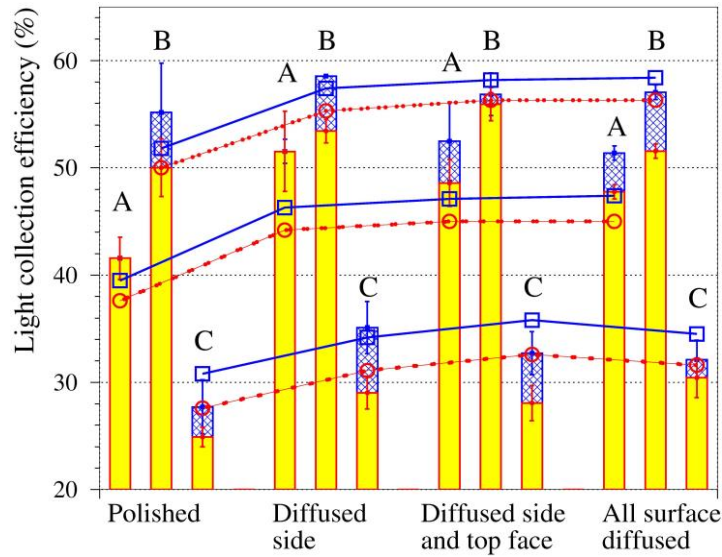

Fig. 6. (Colour online) Relative pulse amplitude for ZnWO$_4$ scintillator crystals (shaded bars for hexagonal and clear bars for cylindrical crystals) for the setups *A, B* and *C* (see Fig. 2) measured with a $^{137}$Cs source for the different conditions of the crystal surface. Light collection efficiency simulated by ZEMAX for the experimental conditions (open squares connected by solid lines for hexagonal, and open circles connected by dashed lines for cylindrical crystals). The experimental data are normalized to the simulated value for the cylindrical polished crystal in the geometry *B*.

## 3. Simulation of scintillation detectors response

There is numerous literature describing Monte-Carlo simulation of light collection in scintillation detectors by using different software (see, e.g. [27, 28, 29, 30, 31, 32, 33] and references therein). Systematic studies of the commonly used reflector materials have also been performed with the aim to develop accurate Monte Carlo simulation models of scintillation detectors [34, 35]. We have used the Geant4 and ZEMAX packages to simulate the experimental data obtained for $ZnWO_4$ scintillators. Geant4 is an advanced tool for simulation of interaction of ionizing radiation with matter with capability of ray-tracing used to simulate the subsequent stages of this interaction, i.e. propagation of light in a scintillation detector. ZEMAX is software developed for modelling of optical systems. In the context of this study it is used exclusively for simulation of light propagation and therefore relies on the prior knowledge on the initial distribution of photons in the detector. However, this slight deficiency is compensated by the flexibility of the software, i.e. the ability to incorporate in a model a whole suite of optical parameters that describe the system under study and then yields adequate results.

It is worthwhile noting that in our simulations of light collection we introduced an additional parameter, the scattering coefficient. The implementation of this parameter leads to the increase of the light output of the crystal as it breaks looping of the photons trapped in the crystal by total internal reflection and therefore it facilitates their escape [17, 30, 31, 36, 37]. However, as there are no methods for accurate direct measurements of the light scattering coefficient in the crystals, this parameter was allowed to vary in the simulations. The value of the light scattering coefficient that gives the best description of the experimental results by simulations was then selected.

### 3.1. Simulation of photoelectron yield by the GEANT4 package

Simulation of the scintillation photon transport with the Geant4 package (v.9.6) had been performed for the geometries described above. The input parameters for the simulations are summarized in Table 1. A source, emitting 662 keV gamma quanta, was positioned on the z axis, far away below the $ZnWO_4$ crystal so that the flux of incident gamma quanta can be approximated by a uniform, parallel beam. The points of light emission (vertices) were generated according to an exponential distribution along the vertical *z* axis of the set-up, characterized by the attenuation coefficient for 662 keV gamma quanta in $ZnWO_4$ (14.47 mm). The distribution of the vertices in the horizontal *xy* plane of the scintillator was uniform.

Table 1. Input data for Monte Carlo simulations.

| Parameter | Geant4 | ZEMAX |
|---|---|---|
| Absorption of γ quanta in $ZnWO_4$ crystal | Calculated by Geant4 | Uniform |
| Emission spectrum of $ZnWO_4$ | [38] | 480 nm (max) |
| Absorption length of the $ZnWO_4$ crystals | 24 cm (at 480 nm) | 24 cm |
| Diffuse crystal surface | 50% Lambertian + 50% specular | Lambertian |
| Reflectivity properties of Teflon tape | [34] | 100% |
| Reflectivity properties of 3M reflector | [34] | 100% |
| Transmittance of optical contact material | Data of producer | 100% |
| Bulk scattering coefficient | 0.065 $cm^{-1}$ | 0.2 $cm^{-1}$ |
| Anisotropy of $ZnWO_4$ | No | No |
| Quantum efficiency of the PMT photocathode | Data of producer | No |

The optical properties of all media were taken to be isotropic. An index of refraction for $ZnWO_4$ of 2.3 was used. We assume as a good approximation that the indices of refraction of all media are constant over the scintillator emission range (380 – 700 nm). The scintillation spectrum has been taken from [38]. The photons in the vertex were generated with a space-isotropic distribution of momentum and plane-isotropic polarization orthogonal to momentum. The energies of photons were generated in agreement with the emission spectrum multiplied by the spectral sensitivity of the PMT, in order to avoid spending processor time for tracking the photons which will not produce photoelectrons. As the spectral sensitivity of the photocathode is taken into account at the time of photon generation this allows to consider the PMT photocathode as being a 100% efficient photon detector. The number of generated photons was defined by the emission spectrum alone.

The absorption spectra of $ZnWO_4$ were normalized such that the absorption length of the cylindrical crystal was equal to 24 cm at the wavelength of the emission maxium. Bulk scattering was implemented with Rayleigh wavelength dependence of $1/\lambda^4$ and normalized in such way that the scattering coefficient at 500 nm was equal 0.065 $cm^{-1}$.

The ground surfaces of the crystals were assumed to be 50% polished and 50% diffuse (Lambertian). An assumption regarding imperfection of the diffuse crystal surface was confirmed by micrograph of the crystal surface as shown in Fig. 3. The diffuse surface of the crystals still exhibits partly specular reflection. The optical properties of 3M film and teflon were taken from the RealSurface1.0 data set [34] of Geant4 (PolishedVM2000 and GroundTeflon, respectively). We simulated 1000 vertices with 1000 photons emitted in each vertex. The results of the simulations are presented in Fig. 7 in comparison with the experimental data normalized to the simulated value for the cylindrical polished crystal in geometry *B*. This experimental condition is deemed to be simulated with the best accuracy. The simulations describe qualitatively the experimental data, though the discrepancy between the theoretical calculations and experiment is still quite noticeable.

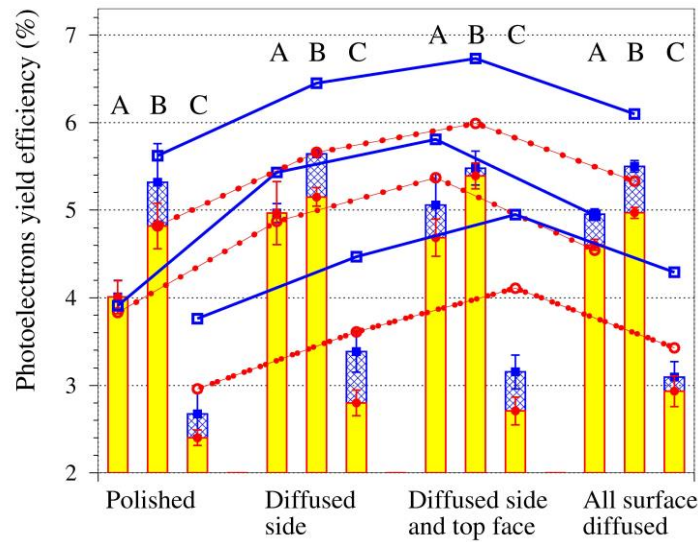

Fig. 7. (Colour online) Photoelectron yield efficiency (ratio of created photoelectrons to emitted photons, %) for different setups and surface conditions of cylindrical and hexagonal $ZnWO_4$ scintillators calculated by the Geant4 package (open squares connected by solid lines for hexagonal, and open circles connected by dashed lines for cylindrical crystals). The experimental data (shaded bars for hexagonal and clear bars for cylindrical crystals) are normalized to the simulated value for the cylindrical polished crystal in geometry *B*.

### 3.2. Simulation of light collection using ZEMAX code

The ray-tracing software ZEMAX was used to simulate light transport in the $ZnWO_4$ scintillators. First, three models were created to reproduce the different experimental setups *A*, *B* and *C* with the scintillation volume in the shape of a hexagonal prism and cylinder. All simulations were carried out in non-sequential mode. It means that there is no predefined sequence of surfaces that traced rays must hit. The trajectories of rays are determined solely by the physical positions and properties of the objects and directions of the rays. Thus, the rays may hit any part of any object, and thus may hit the same object multiple times. This allows accounting for total internal reflections and taking into consideration polarisation effects and ray splitting at the interfaces.

The source was simulated as a volume object of cylindrical or hexagonal shape with dimensions 0.001 mm less than the geometrical sizes of scintillation crystal to eliminate uncertainty at the surface. The wavelength was set to the peak wavelength of $ZnWO_4$ emission at 480 nm. Zinc tungstate is an anisotropic crystal with noticeable variation of refractive index over direction. Comprehensive accounting for this effect significantly complicates modelling but adds very little to the final results. It was estimated that the results varies within ±3% if the index of refraction changes from its maximum $n_a$=2.40 to its minimum value $n_c$=2.23 at 480 nm [39]. Therefore the simulations were done in isotropic approximation using an average dispersion curve and refractive index $n_b$=2.30 at 480 nm. The absorption coefficient of the crystal was obtained from measurements of transmission after correction for multiple reflections following [40]. The entrance window of the PMT was modelled as a 78 mm × 1mm disk made of borosilicate glass BK7 with refractive index 1.52. For the calculation of light collection efficiency the detector was assumed to be an ideal absorber at the inner surface of the PMT window. The coupling of the crystal to the detector using optical gel (setup *A* and *B*) was modelled as a 0.1 mm slice of silicon with the refraction index 1.46. The Teflon tape used in setup *B* was represented as an additional reflector with diffuse (Lambertian) scattering distribution applied to the back and side of the scintillator crystal. In experiment, the condition of different surfaces of the crystals (side, back and front) was subsequently altered by applying lapping. Such treatment changes a surface state from polished to diffuse. To account for this change in the simulation a lapped surface was modelled as a Lambertian diffuser.

In each case, the 100,000 rays randomly distributed across the volume of the crystal were traced. The result of the simulations is the fraction of the total energy generated by the source object (scintillator) that reaches the detector. This number represents the light collection efficiency of the setup. As the bulk scattering coefficient of the crystal is an unknown parameter the ray tracing was carried out for four values: $\alpha_{scat}$ = 0.1, 0.2, 0.5 and 1 cm$^{-1}$. Fig. 6 shows the simulation data obtained for the scattering coefficient $\alpha_{scat}$ = 0.2 cm$^{-1}$ that gives the best agreement with the experimental results.

## 4. Discussion

Within of the three experimental geometries under study, setup *B* exhibits the highest light collection efficiency (above 50%). Setup *C* has the lowest light collection efficiency (ca. 30%) as the air gap between the scintillator and the detector leads to significant light loss. The simulation also reveals that the application of the external specular reflector (setup *B*) is more efficient for light collection than the diffusive Teflon wrapping of the scintillation crystal used in setup *A*. This is due to the fact that in setup *B* the gap between the scintillation crystal and external reflector provides an escape channel for a fraction of photons allowing them to reach detector (see Fig. 8). It is worthwhile to mention important feature observed for polished cylindrical crystal in this geometry: despite of the increase of light collection efficiency the energy resolution significantly degrades as opposed to the geometry A. The reason for this is a significant nonuniformity of light collection that is inherent property of cylinder. The closer a photon is generated near the curved cylindrical surface the higher is the chance for trapping [41]. As a matter of fact, cylindrical crystals behave like a light guide that has a significant radial variation of the light distribution [42] as is shown in Fig.8. This results in degradation of energy resolution of polished cylindrical crystal.

The degradation of energy resolution for a polished cylinder as in geometry C is even more dramatic. This feature has been observed in the CRESST experiment that initially used polished cylinder scintillators [43]. The non-uniformity of light collection across the crystal volume was shown to be a main cause of this effect as demonstrated by [44]. The lapping of the crystal surface allowed to mitigate this problem of CRESST detectors, and that is consistent with our observations for the cylindrical geometry of the scintillator crystal.

For all studied set-ups and experiments, the light collection efficiency of the scintillator in the shape of a hexagonal prism is better than that of a cylindrical scintillator. Obviously, the circular shape is a cause of light trapping inside the crystal. This is not the case for the scintillator with hexagonal shape as photons traveling at any angle to the side surface can escape the crystal after a few internal reflections.

It is found that for all studied set-ups and experiments the light collection efficiency increases by up to 5% when the side surfaces of the scintillation crystal are made diffuse. Such treatment introduces diffuse reflection at surfaces that allows a fraction of the trapped light to be re-directed and escape the crystal. The lapping of the back and front surfaces has little effect.

Juxtaposition of normalised experimental data and the ZEMAX modelling results shows good agreement in most cases (see Fig. 6). The simulated light output of different set-ups after scaling is consistent with the measured results. The simulations slightly underestimate the effect of Teflon tape on the light collection. This mismatch is likely to be due to the complexity in simulation of wrapping under experimental condition [36, 37]. Nonetheless, the main trends discussed above follow experimental observations. This evidences the merit of a simulation model as well as its potential to predict performance of a scintillation detector and provide guidelines for further optimisation.

Interestingly, the simulations performed by using the ZEMAX software describe the experimental data better than that obtained with the Geant4 package, despite the fact that in the Geant4 simulations a more precise model of the scintillation detectors was used (particularly, the absorption of γ quanta in $ZnWO_4$ crystals and spectral properties of the photomultiplier cathode were taken into account). The advantage of ZEMAX over another Monte Carlo simulation package, Detect 2000, when applied for simulating light collection has already been noticed previously [32].

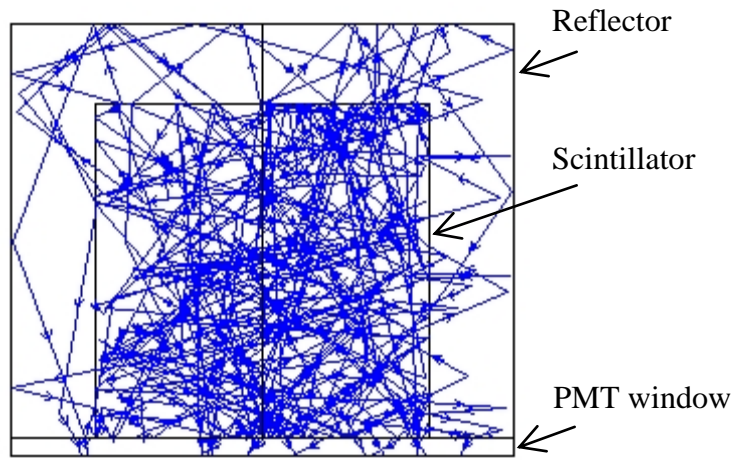

a)

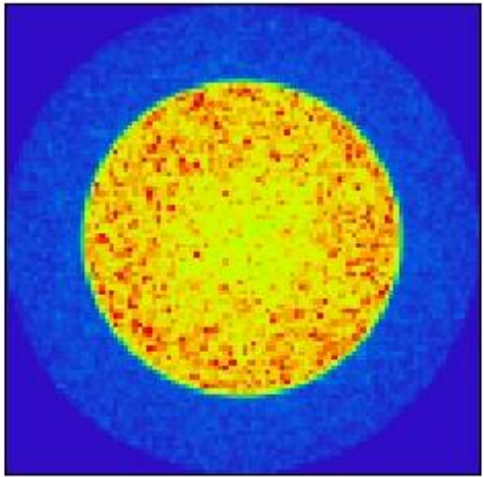

b)

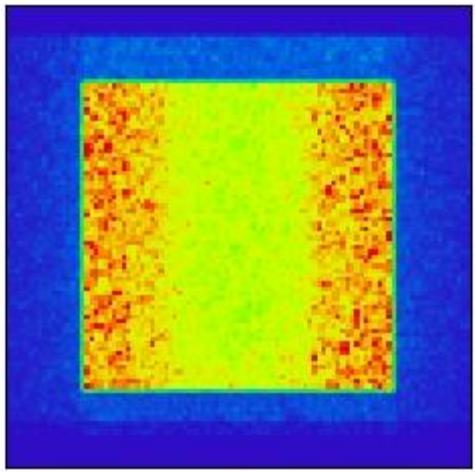

c)

Fig. 8. (Colour online) Trajectories of optical photons simulated by ZEMAX for cylindrical ZnWO$_4$ scintillation detector in geometry *B* (a) and distribution of illumination in the horizontal and vertical planes of the cylindrical crystal (b) and (c), light intensity is represented by colour increasing from blue to red (see on-line version). The images illustrate the radial distribution of light inside of the crystal: more trapped photons are traveling closer to the curved cylindrical surfaces. The images also show how an external mirror surface redirects a fraction of the light through the gap between the scintillation crystal and the reflector.

Our measurements were carried out at room temperature with a PMT as photodetector. There are a few issues, which should be taken into account in further optimization of low temperature scintillating bolometers: 1) the light response of a scintillating crystal generally increases at cryogenic temperature; 2) germanium or silicon slabs used as photodetectors at present have substantially different optical properties in comparison to a PMT window with photocathode however studies are in progress for the implementation of PMT at cryogenic temperatures as a future option [45]; 3) construction of low temperature scintillating bolometers is slightly different from our simple model; it includes crystal and photodetector holders, sensors, wires typically unnecessary in a room temperature scintillation detector assembly. The exact effect of these differences can be assessed only by carrying out experiments at cryogenic temperatures. However, with respect to light collection efficiency the scintillator in the shape of hexagonal prism has better performance than the cylindrical shape. Besides, the scintillators with diffused surface provide higher light collection and better energy resolution in comparison to polished.

## 5. Conclusions

In this study we investigated the influence of crystal shape, reflector, optical contact and surface treatment on the scintillation characteristics of $ZnWO_4$ scintillators. Two samples of crystal scintillators in the shape of hexagonal prism (height 20 mm, diagonal 20 mm) and cylinder (height 20 mm, diameter 20 mm) were produced from the same crystal ingot. The scintillation characteristics of the crystals were studied for different experimental conditions. The hexagonal crystal shows better energy resolution and relative pulse amplitude for all tested conditions. The best energy resolution (FWHM = 9.2% for 662 keV γ quanta of $^{137}Cs$) was obtained for the hexagonal scintillator with all surfaces diffuse, in optical contact with the PMT, and surrounded by a cylindrical reflector (3M). The maximal light output was measured for the hexagonal scintillator with side surfaces diffuse and polished end faces, in optical contact with the PMT and surrounded by a cylindrical reflector (3M) with a gap between the scintillator and the reflector.

In the geometry "without optical contact" representing the conditions of light collection in cryogenic scintillating bolometer the light collection efficiency is reduced by a factor of two when compared with the other experimental geometries. This is due to substantial reflection loss at the interfaces between the crystal and light detector. The energy resolution of the detector, though degraded, remains useful for practical applications. In this geometry the best relative pulse amplitude and energy resolution (FWHM = 11.8 % for 662 keV γ quanta of $^{137}Cs$) were obtained for the hexagonal shape scintillator with diffuse side surface.

Despite the measurements were at room temperature they provide directions to improve the light collection in cryogenic experiments. Besides, the measurements are useful to develop methods of Monte Carlo simulations of photons propagation in scintillators independent of the operating temperature. The responses of the scintillation detectors were simulated using Geant4 and the ZEMAX software. We succeeded to adjust the ZEMAX model to the experimental data, the largest error between simulation and experiments being about 10%. At the same time, the results obtained with Geant4 demonstrate a more significant deviation (up to 50%), though qualitatively they reproduce general trends in the characteristics of scintillation light output of detector correctly.

**Acknowledgements**

The study was supported in part by a grant from the Royal Society (London) ''Cryogenic scintillating bolometers for priority experiments in particle physics''. The group from the Institute for Nuclear Research (Kyiv, Ukraine) was supported in part by the Space Research Program of the National Academy of Sciences of Ukraine. V.V.K. was supported by the